\documentstyle[aps,epsf]{revtex}
\def\be{\begin{equation}}
\def\ee{\end{equation}}
\def\bea{\begin{eqnarray}}
\def\eea{\end{eqnarray}}

\begin{document}
\title{Initial data and spherical dust collapse}
\author{Filipe C. Mena$^1$, Reza Tavakol$^1$ \& Pankaj S. Joshi$^2$}
\date{November, 1999}
\maketitle
{\em $^1$ Astronomy Unit, 
School of Mathematical Sciences,  
Queen Mary \& Westfield College, 
Mile End Road, 
London, E1 4NS, U.K.}
\\

{\em $^2$ Tata Institute of Fundamental Research, Homi Bhabha Road, Bombay 400 005, India}
\\

\begin{abstract}
By considering families of radial null geodesics, we study the subsets of initial data
that lead to naked singularities and black
holes  
in inhomogeneous spherical dust collapse.
We introduce the notion of {\it central homogeneity}
for spherical dust collapse and prove that for the occurrence of 
naked singularities, the initial data set
must in general be centrally homogeneous.
Even though mathematically this indicates
that naked singularities are in general unstable, we
show that centrally inhomogeneous
perturbations in the initial data are not physically reasonable. 
This provides an example of the fact
that instability in this context deduced with respect to
general perturbations can become stabilised
once the class of perturbations are 
restricted to be physical.

This is a potentially important point to bear in mind
in the general debates regarding the generic presence
of naked singularities in gravitational collapse and
in more general debates concerning the questions
of genericity in general relativity.
\end{abstract}
\section{Introduction}
It is now well known that, subject to
a number of physically reasonable assumptions,
final state singularities arise
in solutions of Einstein's equations in a
range of settings (see e.g. \cite{Hawking-Ellis73,Wald97}).
What is not known in general - and this is
one of the key outstanding questions in 
classical general relativity at the moment -
is the nature of the resulting singularities
and in particular whether and under what conditions
they may be naked (NS) or black holes (BH) 
\cite{Penrose,Joshi-Book}.

A great deal of effort has gone
into the study of this question
over the recent years. Given the complexity of
the full Einstein's equations, these studies have
mainly concentrated on the spherically symmetric collapse and 
fall into a number of categories.
One of the mathematically most developed of these, 
due to Christodoulou \cite{chris94,chris99},
concentrates on the spherical gravitational collapse 
of a scalar field and shows that naked singularities
do indeed occur \cite{chris94}, but that in a certain sense 
they are unstable \cite{chris99}.

Another group of works has concentrated on 
showing the occurrence of NS solutions in a variety of spherical symmetric
space--times with several examples of field--sources,
including dust \cite{Dwivedi-Joshi97}, perfect fluids \cite{Ori-Piran90,Joshi-Dwivedi92},
imperfect fluids \cite{Magli97,Lake92} 
and radiation \cite{Joshi-Book}.
In particular it has been shown that in spherical dust collapse,
given any initial 
density profile for the collapsing cloud,
the corresponding velocity profile may be chosen such that
the collapse may eventually result either in a BH or a NS
\cite{Dwivedi-Joshi97}.

In this way, both groups of works demonstrate that
the end result of spherical collapse 
depends upon the
nature of the initial data. 
\\

Our aim here is to make a more thorough study
of the outcomes of the inhomogeneous spherical
dust collapse and
in particular to study the nature of
subsets of initial data ${\cal I}$
corresponding to NS and BH.
\\

The structure of the paper is as follows.
In section 2 we give a brief description
of the spherical dust collapse. In Section
3 we summarise physical and other constraints
that need to be satisfied by functions
that represent the initial data in these models.
Section 4 contains our main results in the
form of a number of Lemmas and Propositions
and finally Section
5 contains our conclusions.
\section{Spherically symmetric dust collapse}
The inhomogeneous spherically symmetric dust collapse can
be represented by the Lemaitre-Tolman-Bondi (LTB)
line element which 
is given by \cite{Lemaitre,Tolman,Bondi}
\be
\label{metric-tolman} 
ds^2 = -dt^2 + \frac{{R^{'}}^2}{1+E} dr^2 +R^2 (d\theta^2 + \sin^2
\theta
d\phi^2),
\ee
where $r, \theta, \phi$ are the comoving coordinates. 
The dot and prime denote differentiation with
respect to $t$ and $r$ respectively and
$R=R (r,t)$ and $E=E(r)$ are $C^2$ real functions
such that $R (r,t)\ge 0$ and $E(r)>-1$. 
The matter--density is given by
\be
\label{density}
\rho(t,r) = \frac{M^{'}}{R^2 R^{'}}
\ee
where $M=M(r)$ is another $C^2$ real function such that $M(r)>0$. 
The evolution equation for the case of $\dot{R}<0$
(corresponding to gravitational collapse) 
takes the form
\be
\dot R = - \sqrt{\frac{M}{R} + E}
\ee
which can be solved to give
\be
\label{evolution}
t-t_c(r)=-\frac{R^{3/2}G(-ER/M)}{\sqrt{M}},
\ee
where $t_c=t_c(r)$ is another real $C^2$ function that corresponds to 
the time of arrival of each shell $r$
to the central singularity and 
$G$ is a positive function given by
\bea
\label{Gfunctions}
&G(x)=&\frac{arcsin\sqrt{x}}{x^{3/2}}-\frac{\sqrt{1-x}}{x},~~~~~~~~
for~~~1\ge x>0\nonumber\\
&G(x)=&2/3,~~~~~~~~~~~~~~~~~~~~~~~~~~~~~~~ for~~~ x=0\\
&G(x)=&\frac{-arcsinh\sqrt{-x}}{(-x)^{3/2}}-\frac{\sqrt{1-x}}{x},
~for~~~ 0> x > -\infty.
\nonumber
\eea
Using the coordinate freedom to rescale 
\be
R(0,r)=r,
\ee 
equation (\ref{evolution}) gives 
\be
\label{tnot}
t_c(r)=\frac{r^{3/2}G(p)}{\sqrt{M}},
\ee
where $p=-r\frac{E}{M}$. The collapse is then
simultaneous for all shells if
$t'_c(r)=0$, which is the case for the homogeneous
dust collapse.

We note that the metric (1) can be matched at the boundary, say $r=r_d=const.$,
to the Schwarzschild metric in the exterior region \cite{Krasinski}. Thus the scenario here
is that of a collapsing compact matter region
matched in the exterior to the Schwarzschild geometry.

We shall refer to a singularity as
naked if there is a family of future directed non--spacelike
geodesics which terminate at the singularity in the past.  Here we consider the 
outgoing radial null geodesics which, as can be seen from (\ref{metric-tolman}), 
correspond to the solution of the differential equation
\be
\label{geo}
\frac{dt}{dr}=\frac{R'}{\sqrt{1+E}}.
\ee
One can now rewrite this equation as
\be
\label{geod}
\frac{dR}{du}=\frac{1}{u'}\left(R'+\dot{R}\frac{dt}{dr}\right)=\left(1-
\sqrt{\frac{E+\Lambda/X}{1+E}}\right)H(X,u),
\ee
where
\be
H(X,u)=(\eta_u-\beta_u)X+
(\Theta_u-(\eta_u-\frac{3}{2}\beta_u)X^{3/2}G(-PX))\sqrt{P+1/X}
\ee
and
\bea
&&X=\frac{R}{u},~\eta= r\frac{M'}{M},~\eta_u= \eta \frac{u}{ru'},~
\beta=r\frac{E'}{E},~\beta_u=\beta \frac{u}{ru'},\nonumber\\
&&P=\frac{uE}{M},
~p=-r\frac{E}{M},~\Theta_u=\Theta\frac{\sqrt{r}}{\sqrt{u}u'},~
\Lambda=\frac{M}{u},\\
&&\Theta=\frac{\sqrt{M}}{\sqrt{r}}t'_c(r)=
\frac{1+\beta-\eta}{\sqrt{1-p}}+(\eta-\frac{3}{2}\beta)G(p),\nonumber
\eea
with the positive real function $u=u(r)$ being monotonically increasing and such that $u(0)=0$. Later
we will specify $u(r)=r^\alpha,~\alpha>0$.
In the cases where $E(r)=0$ we will take 
$\beta(r)=p(r)=P(r)=0$ by convention. 
 
A subscript $`0`$ will denote the limit of the associated  
functions as $r\to 0$ (respectively $u\to 0$). We should emphasise that
the existence of such limits cannot be assumed a priori and need
to be ensured for the given set of
initial data under consideration, as we shall do 
in the following.
 
It can be shown from \cite{Joshi-Dwivedi93} 
that a sufficient condition for the
existence of a naked singularity 
in spherical symmetric dust collapse 
is that the following algebraic equation in $X_0$
\be
\label{general}
\left(1-\sqrt{\frac{E_0+\Lambda_0 /X_0}{1+E_0}}\right)H(X_0,0)-X_0=0,
\ee 
possesses a real positive
root. These roots give the possible values of 
tangents for the outgoing geodesics at the singularity such that
the associated integral curves terminate at the singularity in the past. 
We note that given the limiting nature of 
(\ref{general}), the forms of 
$E$ and $M$ in a neighbourhood of $r=0$ will
play an important role in determining
the possible solutions to this equation.

An interesting outcome of all the studies of the spherical
dust collapse in the literature
is that for the initial functions $E$ and $M$
chosen so far, the most general
form of the equation (\ref{general})
becomes a polynomial of degree not greater than
four \cite{Joshi-Dwivedi93,Dwivedi-Joshi97}.
As we shall see in the next section, this feature
is very important in constraining the
way the subsets of initial data corresponding to
NS and BH are distributed in ${\cal I}$. As a result it
is important to ask whether a quartic is the 
most general form equation (\ref{general}) can take.

Another important outcome for these studies is that the occurrence of BH or NS as 
final outcomes of collapse depend on the choice of initial data.
The question arises as to the nature of the subsets of the initial data
that lead to each of these outcomes and how robust are these outcomes with 
respect to perturbations in the initial data.
\\

Before discussing these issues in Section 4, 
we briefly consider, in the next Section, some constraints
that are to be satisfied by the functions 
$E$, $M$ (and $\Theta$) on physical grounds.
\section{Physical constraints}
\label{constraints}
In the case of spherical dust collapse
the initial data are given in terms of
two functions; namely the mass function for the dust cloud, $M=M(r)$, and the energy
function $E=E(r)$, which is related to the initial radial velocity $V_I(r)=\dot{R}(0,r)$ 
of the shells by
\be
E(r)=V^2_I(r)-\frac{M(r)}{r}.
\ee
Here we briefly summarise the constraints 
that these functions, as well as $\Theta$,
need to satisfy in order to be
physically acceptable.
To begin with, to ensure that the curvature is initially
well behaved at the regular centre of the matter distribution, 
we demand the condition that
the quadratic curvature scalar
(the so called Kretschmann scalar) given by
\be
\label{Kretchmann}
K=R^{ijkl}R_{ijkl}=\frac{12M'^2}{R^4R'^2}-\frac{32MM'}{R^5R'}+\frac{48M^2}{R^6}
\ee
remains bounded.
We should note here that in this case
the Ricci scalar is the only term that remains
relevant near origin, if the origin is regular.
We have employed the Kretschmann scalar because in addition to encoding all
the information we need from the Riemann invariants, it enables us to ensure that
the centre is regular.

To ensure this, as well as the 
finiteness of the density distribution at the initial
epoch, $M$ in the neighbourhood of the origin needs, in general, to be 
of the form\cite{gen}
\be
\label{Mgeneral}
M(r)=O(r^a),~a\ge 3.
\ee
If we require $\rho(0,0)\ne 0$ then we need in general\cite{rho-non-zero}
\be
\label{M}
M(r)=r^3 g(r)+O(r^m),~m>3,
\ee
where $g$ is a $C^2$ function of $r$ that remains 
finite as $r\to 0$. 
The condition for the absence of shell-crossing is
given by \cite{Dwivedi-Joshi97} 
\be
\Theta(r)\ge 0.
\ee 
We also note that the apparent horizon is given
by $R(r,t)=M(r)$ \cite{Dwivedi-Joshi97} and in order 
to ensure that 
the initial hypersurface does not contain any trapped surfaces
we shall require the condition $M(r)/R(r,0)<1$.
By a {\em regular initial data} we will mean 
initial data that satisfies the physical conditions given above.
\\
 
For simplicity in the following we shall, unless
otherwise stated, 
assume that, in a neighbourhood of $r=0$, 
the functions $E=E(r)$ and $M=M(r)$ can be expressed as\cite{gen2}
\be
\label{poly}
 M(r)=\sum_{i=3}^{\infty} M_{i} r^{i},~~~E(r)=\sum_{j=0}^{\infty} E_{j} r^j,
 ~~~M_i, ~E_j\in \Re.
\ee
We shall refer to the first and second non-vanishing
powers of $M$ and $E$ by  $i_0$, $j_0$ and $i_1$, $j_1$ respectively.
From here on we shall take $u(r)=r^\alpha,~\alpha>0$ \cite{geo-scaling}.
 Letting $R=X_0r^{\alpha}$ in the neighbourhood of the singularity then
$K\propto r^{2(i_0-3\alpha)}$ which implies that $K$ diverges as $r\to 0$ only
along geodesics with
$\alpha>i_0/3$ which for the case $i_0\ge3$ gives $\alpha>1$. 
We shall use this condition in the next
section in the proofs
of the Lemmas.
\\

Finally we recall that for homogeneous
initial data with $E(r)\ne 0$ we have $M(r)^2=k E(r)^3$, $k$ constant,
 which for initial data of the form (\ref{poly}) gives $t_c(r)=const$, 
which in turn implies a simultaneous
collapse \cite{homo}.
The next definition will be useful in what follows:
\\
 
\noindent {\bf Definition:} A LTB initial data set
is said to be
{\it centrally homogeneous} if, in a neighbourhood of the origin, given $E(r)\ne 0$,
$M(r)^2 \propto E(r)^3$ or given
$E(r)=0$, $M(r) \propto R(0,r)^3$.
\\
 
\noindent In particular, a LTB initial data set with $E(r)\ne 0$ is centrally homogeneous
if near the origin                                  
\be       
\label{asymp}
 M(r)=\sum_{i=3}^{\infty} M_{i} r^{i},~~~E(r)=\sum_{j=2}^{\infty} E_{j} r^j;
 ~~~M_3>0, ~E_2\ne 0,
\ee

In the following we shall refer to
perturbations 
which break the central homogeneity of the initial data as
{\em centrally inhomogeneous perturbations}.
\section{Initial data and spherical dust collapse}
In this section we study the final 
outcomes of the spherical 
dust collapse as a 
function of the choice of initial data, by employing a family of outgoing null geodesics
from the origin. In particular, we study the subsets of regular initial data which give
rise to NS solutions in (\ref{general}). As was shown in
\cite{Dwivedi-Joshi97}, finiteness of $\Lambda_0$ is a necessary
condition for the existence of NS solutions. To ensure this we require
that
$i_0\ge \alpha$. We now
note that allowing  $\Theta_{u}, P$ or $p$ to diverge as
$r\to0$ would make equation (\ref{general}) singular at $r=0$, for any regular
initial data with $i_0\ge \alpha$ and $\alpha>1$. Therefore in the following we shall assume that
$\Theta_{u_0}, P_{0}$ and $p_{0}$ are finite.

We begin with a simple (well known)
result which demonstrates that in the homogeneous
setting the set of initial data which lead to
BH is full.
\\

\noindent {\bf Lemma 1}: {\em Consider the spherically symmetric
dust collapse with initial data given by the functions $E$ and $M$
in the form (\ref{poly}) and assumed to be homogeneous.
Then
equation (\ref{general}) has no NS solutions.}
\\

\noindent {\bf Proof:} The homogeneity condition for the functions (\ref{poly})
 implies $2i_0=3j_0$, in the case $E(r)\ne 0$. 
 On the other hand, the requirement of finiteness of $p_0$
 implies $1+j_0-i_0\ge 0$ which results in $i_0\le 3$. From condition (\ref{Kretchmann}) 
 we obtain $i_0=3$ and the homogeneity implies $j_0=2$.
 This gives $\Theta(r)=0$ and the
 equation (\ref{general}) has then the solutions
 \be
  X_0=0~~\vee~~X_0=
 \frac{\sqrt{\Lambda_0}}{1-\alpha},
 \ee
 which for $\alpha>1$ give $X_0\le 0$. The same conclusion holds for the homogeneous 
 marginally bound case where $E(r)=0$ and $i_0=3$~~$\Box$.    
\\

We note that the existence of real positive roots to
equation (8) 
basically characterizes
the formation and time development of the apparent horizon as the 
collapse develops \cite{Dwivedi-Joshi97}.
To understand further the structure of the initial 
data in the inhomogeneous dust collapse,
it is important, as a first step, to establish how general
are the conditions for which 
(\ref{general}) is a polynomial of degree $\le 4$.
The following Proposition makes precise the forms 
that equation
(\ref{general}) may take in order 
for the spherical symmetric dust collapse 
to result in a NS
solution.
\\

\noindent {\bf Proposition 1}: {\em Consider the spherically symmetric
dust collapse with initial data given by the functions $E$ and $M$
in the form (\ref{poly}) and assumed to be regular.
Then the only non-divergent forms of the equation (\ref{general}) 
are polynomials of degree not greater than four.}
\\

\noindent {\bf Proof:} For equation (\ref{general}) to
remain finite in the limit $r\to 0$, the limiting
quantities $P_0,E_0, \Lambda_0$ and $\Theta_{u_0}$
need to remain finite.
Let $x_0=\sqrt{X_0}$  and recall that $u=r^{\alpha}$
and that $j_0$ and $i_0$ are the orders of 
the first non-vanishing coefficients of $E$ and $M$
respectively.
We proceed by considering all different combinations
of these limiting quantities in turn.
\\

\noindent (i) Let $P_0=E_0=0$, $\Theta_{u_0}\ne 0$ and $\Lambda_0\ne 0$.
Then equation (\ref{general}) becomes
\be
\label{quartic}
\left(\frac{1}{3}\eta_{u_0}-1 \right) x_0^4-\frac{1}{3}\sqrt{\Lambda_0}
\eta_{u_0}x_0^3+\Theta_{u_0} (x_0-\sqrt{\Lambda_0})=0.
\ee
The exponent $\alpha$ can then be chosen such that $\Theta_u,\eta_u$ 
and $\Lambda$
remain finite at $r=0$.
\\

\noindent (ii) For $P_0=E_0=0$, $\Theta_{u_0}\ne 0$ and $\Lambda_0= 0$, equation
(\ref{quartic}) gives 
\be
\label{linear}
x_0=0 \vee \left(x_0^3=\frac{\Theta_{u_0}}{1-\eta_{u_0}/3}
\wedge \eta_{u_0}\ne 3\right).
\ee

\noindent (iii) For $P_0=E_0=0$, $\Theta_{u_0}=0$ and $\Lambda_0=0$,
equation (\ref{quartic})
reduces to
\be
\label{universal}
\left(\frac{1}{3}\eta_{u_0}-1 \right) x_0^4=0,
\ee
which has the solutions $x_0=0$ or $\frac{1}{3}\eta_{u_0}-1=0$. 
\\

\noindent (iv) For $P_0=E_0=0$, $\Theta_{u_0}= 0$ and $\Lambda_0\ne0$, equation
(\ref{quartic}) gives 
\be
\label{four}
x_0=0 \vee \left(x_0=\frac{\sqrt{\Lambda_0}\eta_{u_0}}{\eta_{u_0}-3}\wedge
\eta_{u_0}\ne3\right).
\ee

\noindent (v) Let $P_0\ne 0$, $E_0=0$, $\Theta_{u_0}\ne 0$ for any finite $\Lambda_0$.
We first consider $p_0\ne 0$ (i.e., $1+j_0-i_0=0$) which, since $P_0 \ne 0$ (i.e., 
$\alpha +j_0-i_0=0$), implies $\alpha=1$ and 
$1+\beta_0-\eta_0=0$. Requiring $i_0\ge 3$ gives $\eta_0-\frac{3}{2}\beta_0\le 0$
which either contradicts the assumption $P_0\ne 0$
or violates the shell-crossing condition. 
Suppose now $p_0=0$ which implies $\alpha<1$ and in turn $\Theta_{u_0}=0$, contradicting
the assumption $\Theta_{u_0}\ne 0$ and resulting
in equation (\ref{general}) to become singular.
\\

\noindent (vi) Let $E_0\ne 0$ for any finite values of $P_0, \Theta_{u_0}$ and $\Lambda_0$. Then
the only way to make $E_0\ne 0$ (i.e., $j_0=0$) and $p_0$ finite (i.e., $1+j_0-i_0\ge 0$) 
is to assume $i_0\le 1$, which makes 
(\ref{Kretchmann}) diverge and thus results in equation (\ref{general})
to become singular.  
\\
 
\noindent (vii) Let 
$E_0=\Theta_{u_0}=\Lambda_0=0$ and $P_0\ne 0$. Then from (v) above, with $p_0\ne 0$,
we obtain $1+\beta_0-\eta_0=0$ and $\alpha=1$. 
Now since $\Theta_{u_0}=0$, we necessarily have 
$\eta_0-\frac{3}{2}\beta_0= 0$ which gives $i_0=3$ and $j_0=2$
and equation (\ref{general}) becomes identically satisfied for all $x_0$. 
The other
possibility is $p_0=0$ which implies $\alpha<1$ which in general
results in

\be
\label{strange}
x_0=0 \vee \left [
(\eta_{u_0}-\beta_{u_0}-1)-x_0(\eta_{u_0}-\frac{3}{2}\beta_{u_0})
G(P_0 x^2_0)\sqrt{P_0+\frac{1}{x^2_0}}=0 \right ].
\ee
Now since $P_0\ne 0$ implies  $\alpha+j_0-i_0=0$, the first term in
the square bracket 
in the above equation vanishes. On the other hand, the second term would vanish if
$x_0=0$ or $\eta_0-\frac{3}{2}\beta_0=0$ (i.e. $i_0-\frac{3}{2}j_0=0$) which,
since $p_0=0$, implies $i_0<3$, hence making $K$ diverge initially.  
The same term can also vanish if
\be
x^2_0=-\frac{1}{P_0},
\ee
for any $P_0<0$.$~~~~$$\Box$
\\

This Proposition shows that in the case of spherical dust collapse, equation 
(\ref{general}) can only take a restrict number of forms 
with the most general being a polynomial of degree 4 (corresponding to $E_0=0$,
$\Theta_0\ne 0$, $P_0=0$ and $\Lambda_0\ne0$).
\\

Now an important question is the
way the outcome of the spherically symmetric dust collapse 
relates to the central homogeneity or otherwise of the
initial data. The following Proposition makes this 
precise.
\\

\noindent {\bf Proposition 2}: {\em Consider the spherically symmetric
dust collapse with initial data given by the functions $E$ and $M$
in the form (\ref{poly}) and assumed to be regular. Let $p_0 \ne 0$. 
For the occurrence of NS solutions in (\ref{general})
the initial data set must be centrally homogeneous.}
\\

\noindent {\bf Proof:} 
We shall show that if $E$ and $M$ do not satisfy
(\ref{asymp})\footnote{In fact all that is
required is that in the case of $p_0 \neq 0$
the power expansion of 
$E$ and $M$ have lower powers of $r$
given by 2 and 3 respectively.} for $p_0 \ne 0$,
then equations (\ref{quartic})--(\ref{strange}) have no real positive roots.
We shall proceed by considering the following cases:
\\

\noindent (i) Let $p_0=0$, $\alpha<1$. In this case $\Lambda_0=0$,
$\Theta_{u_0}=0$ and a non--zero solution of (\ref{universal})
requires $\eta_0<3$ (ie, $i_0<3$) which
makes $K$ divergent. Case (vi) of Proposition 1 
gives no NS solutions either. In the case of (vii), however, we may find 
NS solutions for $P_0<0$ and $3>j_0>2$ and $i_0=3$.
\\

\noindent (ii) Let $p_0=0$, $\alpha=1$. This implies $P_0=0$ and $\Lambda_0=0$ which,
by requiring $\Theta_0>0$ in (\ref{linear}), means that we would need to have
$\eta_{u_0}=\eta_0<3$, which does not satisfy the condition on $K$. However,
NS may arise in the case (iii) of Proposition 1 with $i_0=3$ and $j_0>2$. 
\\

\noindent (iii) Let $p_0=0$, $\alpha>1$. In this case
\be
\label{Theta}
\Theta_u=(1-\eta/3)r^{\frac{3}{2}(1-\alpha)},
\ee
which implies that in order to make $\Theta_{u_0}$ finite we require
$i_0=3$. But since $p_0=0$
(ie, $1+j_0-i_0>0$) then in addition to $P_0=0$ we also require 
$j_0>2$, which demonstrates that in
this case the presence of NS does not require centrally
homogeneous initial data.
\\

\noindent (iv) Let $p_0\ne 0$, $\alpha<1$. From $p_0\ne 0$ we have $1+j_0-i_0=0$. In the case
of $P_0=0$ 
and $\alpha<1$ we must have $1+j_0-i_0>0$ which is a contradiction.
On the other hand, if $P_0\ne 0$ and $p_0\ne 0$, then $\alpha=1$ which is contrary
to our assumptions and hence there are no NS solutions in this case.
\\

\noindent (v) Let $p_0\ne 0$, $\alpha=1$. These conditions imply $P_0\ne 0$ and therefore
from the case (vii) of Proposition 1 we may have NS solutions only if $i_0=3$ and $j_0=2$.
\\

\noindent (vi) Let $p_0\ne 0$, $\alpha>1$. As in case (iii) above, $\Theta$ must
vanish as $r \to 0$. If we assume $K\ne 0$ initially we need $i_0=3$, but
since $p_0\ne 0$ (i.e., $1+j_0-i_0=0$), then $j_0=2$. 
On the other hand, if we take $K=0$ initially
then from $1+j_0-i_0=0$ we obtain $1+\beta_0-\eta_0=0$ which in order to make $\Theta_{u_0}$
finite necessitates $\eta_0-\frac{3}{2}\beta_0=0$, which implies $i_0-\frac{3}{2}j_0=0$ and
hence $j_0=2$ and $i_0=3$. 
\\

The above considerations show that for $E(r)\ne 0$ and $p_0\ne 0$ we only obtain a NS solution if the lowest order
of the powers of $r$ in $M$ and $E$ are $i_0=3$ and
$j_0=2$ respectively.
\\

In the case of $E(r)=0$ we have $\beta(r)=p(r)=P(r)=0$ and $\Theta(r)=1-\frac{1}{3}\eta$. 
If $\eta_0>3$, then $\Theta<0$ which violates the shell-crossing condition. 
If $\eta_0<3$, then $i_0<3$ which makes $K$ divergent. So, the
only possibility is to have $\eta_0=3$ (ie, $i_0=3$) $~~~$$\Box$.
\\
 
We note that this result
essentially follows from the non-shell-crossing conditions and the initial
regularity of $K$.
\\

The above Proposition shows that unless
$p_0=0$, inhomogeneous neighbourhoods of the
centrally homogeneous initial data
do not in general result in NS. 
Now $p_0=0$
implies that
as we approach the origin the solution describing the matter distribution tends to
a marginally bound solution (given
by (\ref{evolution}) and (\ref{Gfunctions})),
which is rather special. 
This then indicates that
in general in order
to have NS only 
small departures from homogeneity
can be allowed near the origin.
In this sense, NS in spherical dust collapse
may be said to be mathematically
unstable. 

It is, however, important to note that even though
NS may be unstable with respect to general
perturbations (in this case centrally 
inhomogeneous perturbations), they may nevertheless
stabilise if restricted classes of perturbations
are allowed. In particular this seems to
be the case if only physically motivated
perturbations are allowed.
To see this more precisely, recall that
given the forms (\ref{asymp}) for the functions
$M$ and $E$, then the only way to
break the central homogeneity of (\ref{asymp})
is by letting $M_3=0$ or $E_2 =0$.
It turns out, however, that $M_3=0$
would result in a
density profile given by $\rho(r)= M'(r)/r^2$
that tends to zero at the center $r=0$; 
a result contrary to the physical expectation 
of the density being higher at the centre\footnote{In fact the condition for
$\rho'(0,r)>0$ is $rM''(r)<2M'(r)$.}.
Similarly, letting  $E_2=0$, implies
$M_3 =0$ for $p_0 \ne 0$ which would again
result in $\rho$ to become zero at the centre.
 
This gives an important demonstration of the fact
that instability deduced with respect to
general perturbations can become stabilised
once the class of perturbations are restricted
(in this case to physically motivated ones).
\\ 
 
Having demonstrated the relation between NS and 
central homogeneity, it is of importance to be
able to 
determine the subset of
the initial data 
that result
in NS solutions.
The following Lemma makes this precise.
\\

\noindent {\bf Lemma 2}: {\em Consider the spherically symmetric
dust collapse with centrally homogeneous initial data given by the functions $E$ and $M$
in the form (\ref{asymp}) and assumed to be regular. Then the
set of initial data corresponding to NS as
final states possess open intervals in $M_i$ and $E_j$.}
\\
 
\noindent {\bf Proof}:
Considering the case of $\Lambda_0=0$ with $\alpha=5/3$ gives 
\be
\Theta_{u_0}=3\frac{E_3}{E_2}\left(\frac{1}{\sqrt{1-p_0}}-\frac{3}{2}G(p_0)\right)
+4\frac{M_4}{M_3}\left(G(p_0)-\frac{1}{\sqrt{1-p_0}}\right).
\ee
From the case (ii) of Proposition 1 we find that the occurrence of a NS solution
necessitates $\Theta_{u_0}>0$. We also know that 
$G(p)-\frac{1}{\sqrt{1-p}}<0$ for $1\ge p>-\infty$ (see (\ref{Gfunctions}))
and that $\frac{1}{\sqrt{1-p}}-\frac{3}{2}G(p)>0$ for $1\ge p>0$ and
$\frac{1}{\sqrt{1-p}}-\frac{3}{2}G(p)<0$ for $0>p>-\infty$.
Therefore the condition $\Theta_{u_0}>0$ is always satisfied for 
the intervals
\be
M_3\in]1,+\infty[,~M_4\in]-\infty,0[,~E_2\in ]0,1[,~E_3\in]-\infty,0[,
\ee
or
\be
M_3\in]0,+\infty[,~M_4\in]-\infty,0[,~E_2\in ]-\infty,0[,~E_3\in]-\infty,0[,
\ee     
with all other coefficients $M_i$ and $E_j$ (with $j\ge 2$) in (\ref{asymp})
being real and arbitrary.
In this way we have found, for both $E_2<0$ and $E_2>0$, open intervals in all coefficients
$M_i,~ i\ge3$ and $E_j,~ j\ge 2$ such that the final state of collapse is a NS.~~~$\Box$
\\ 
 
As an example of this result, Figure 1 depicts the 
outcomes
of the dust collapse as a function of
the two parameters                                    
$M_3$ and $E_5$, 
(while                                                 
for the sake clarity the other coefficients $M_i$
and $E_j$ in the initial data are kept fixed). 
As can be seen, there exists open neighbourhoods  
of the initial data (in the $M_3-E_5$ plane) resulting 
in each outcome.

\begin{figure}
\centerline{\def\epsfsize#1#2{0.5#1}\epsffile{graph2.eps}}
\caption{\label{graph1}
Final state of collapse as a function of 
the coefficients $M_3$ and $E_5$
of the functions
$M=\sum_{i=3}^{\infty} M_{i} r^{i}$ and
$E(r)=-E_2r^2-\sum_{j=5}^{\infty} E_{j} r^j$, with $E_2$
and all other coefficients kept fixed. The graininess in this 
figure is due to the finite resolution of the mesh of
initial points chosen numerically.}
\end{figure}

The definition of central homogeneity given in
Section 3 constraints the
functions $E$ and $M$ by imposing 
lower bounds on the degrees
of the polynomials in the expansions of these 
functions around the origin. It turns out that for the presence 
of NS, central homogeneity
also places upper bounds on
the second non-vanishing coefficients of $E$ and $M$.
This is due to the fact that the function $\Theta$ 
vanishes at the lowest order for $r=0$, so in order
to ensure the finiteness of $\Theta_{u_0}$ we will have 
to analyse the second non-vanishing terms in $E$ and $M$
and fine tune them in line with the choice of  
$\alpha$ \cite{second-order}. To begin with,  we note that for a 
centrally homogeneous initial data set we have
\be
\Theta_{u_0}=\lim_{r\to 0}\left[j_1\frac{E_{j_1}}{E_2}r^{j_1-2}
\left(\frac{1}{\sqrt{1-p}}-\frac{3}{2}G(p)\right)+i_1\frac{M_{i_1}}{M_3}r^{i_1-3}\left(G(p)-
\frac{1}{\sqrt{1-p}}\right)\right]
r^{\frac{3}{2}(1-\alpha)}.
\ee
The necessary and sufficient condition for 
$\Theta_{u_0}$ to be positive and finite is given by

\bea
\label{theta1}
&\left(j_1=\frac{3}{2}(\alpha-1)+2~ \wedge i_1>\frac{3}{2}(\alpha-1)+3
~ \wedge~ sgn(E_{j_1})=-sgn(E_2)\right)\vee\nonumber\\
&\vee\left(j_1>\frac{3}{2}(\alpha-1)+2~ \wedge
i_1=\frac{3}{2}(\alpha-1)+3~ \wedge
M_{i_1}<0 \right)\vee\\
&\vee\left(j_1=\frac{3}{2}(\alpha-1)+2~ \wedge
i_1=\frac{3}{2}(\alpha-1)+3~ \wedge~ j_1\frac{E_{j_1}}{E_2}
\left(\frac{1}{\sqrt{1-p_0}}-\frac{3}{2}G(p_0)\right)+
i_1\frac{M_{i_1}}{M_3}\left(G(p_0)-\frac{1}{\sqrt{1-p_0}}\right)>0 \right),
\nonumber
\eea
in the case $E(r)\ne0$ and  
\be 
\label{theta2} 
i_1=\frac{3}{2}(\alpha-1)+3~ \wedge
M_{i_1}<0 
\ee
in the case $E(r)=0$, where $sgn$ denotes the sign function.  
\\

Now recall that in the case $i_0=3$ in order to ensure that
the geodesics originating at the singular point $r=0$
can come out and, at the same time, to have $K$
divergent as $r\to 0$, we need to consider outgoing radial null 
geodesics such that 
$\alpha>1$ (see Section 3).
We note also that
the cases in Proposition 1 where real positive solutions 
with $\alpha>1$ may occur in (\ref{general}) correspond to $\Theta_{u_0}>0$, 
i.e. cases (i) and (ii). Cases (iii), (iv) and (vii)
require $\alpha\le 1$ and will not be considered 
in what follows.  
Given these definitions and notations on $i_1$ and $j_1$,
necessary conditions
for the existence of a NS solution in the case of centrally homogeneous initial data
 are made precise through the following
Lemma.
\\
 
\noindent {\bf Lemma 3}: {\em Consider the spherically symmetric
dust collapse with centrally homogeneous initial data given by the functions $E$ and $M$
in the form (\ref{asymp}) and assumed to be regular.
Then in order for (\ref{general}) to have NS solutions    
it is necessary to have}

\bea
&\left(j_1\in(2,5] \wedge i_1>3 \wedge sgn(E_{j_1})=-sgn(E_2)\right)
\vee\left(j_1>2 \wedge i_1\in (3,6] \wedge M_{i_1}<0\right)\vee\nonumber\\
&\vee\left(j_1\in(2,5] \wedge i_1\in (3,6]~\wedge 
~j_1\frac{E_{j_1}}{E_2}
\left(\frac{1}{\sqrt{1-p_0}}-\frac{3}{2}G(p_0)\right)+
i_1\frac{M_{i_1}}{M_3}\left(G(p_0)-\frac{1}{\sqrt{1-p_0}}\right)>0
\right)\nonumber,
\eea
if $E(r)\ne0$ and
\be
i_1\in(3,6] \wedge M_{i_1}<0,
\ee
if $E(r)=0$.
\\

\noindent{\bf Proof:} In order to have a finite value of $\Lambda_0$ 
for centrally homogeneous initial data we 
need $\alpha\le 3$. Since for $\alpha>1$ we need $\Theta_{u_0}>0$ 
to have NS solutions to (\ref{general}) then 
the necessary conditions follow directly from
(\ref{theta1}) and (\ref{theta2}).~~~$\Box$
\\ 

Finally, for the case of the spherical dust collapse we
give necessary and sufficient conditions
on the functions $M$ and $E$, given by (\ref{poly}) near the origin,
for the existence of NS solution to (\ref{general}) in the case $p_0\ne 0$.
The results that follow generalise previous results given 
in \cite{Jhingan-Joshi97} where sufficient conditions for the occurrence of NS 
were obtained considering the functional forms (\ref{asymp}).
\\

\noindent {\bf Lemma 4}: 
{\em Consider the spherically symmetric
dust collapse with initial data given by the functions $E$ and $M$
in the form (\ref{poly}) and assumed to be regular. Let $p_0\ne 0$ and $\Lambda_0=0$.
Then 
there are NS solutions to (\ref{general}) if and 
only if $\Theta_{u_0}>0$ and 
the initial data is centrally homogeneous.}
\\

\noindent{\bf Proof}: The necessary condition of central homogeneity was  
already established in Proposition 2. Now, the case (ii) of Proposition 1 demonstrates
that for centrally homogeneous initial data and $\alpha>1$ equation
(\ref{linear}) has positive real solutions if and only if $\Theta_{u_0}>0$.
~~~$\Box$
\\

\noindent{\bf Lemma 5:} {\em Consider the spherically symmetric
dust collapse with initial data given by the functions $E$ and $M$
in the form (\ref{poly}) and assumed to be regular. Let $p_0\ne 0$ and $\Lambda_0\ne 0$.
Then there are NS solutions to (\ref{general}) if and 
only if }
\be
\Theta_{u_0}\in\left]0,M_3^\frac{3}{2}\left(\frac{13}{3}-\frac{5\sqrt{3}}{2}\right)\right[\cup
\left]M_3^\frac{3}{2}\left(\frac{13}{3}+\frac{5\sqrt{3}}{2}\right),+\infty\right[
\ee
{\em and the initial data is centrally homogeneous.}
\\

\noindent{\bf Proof:} From Proposition 2 we know that the initial data
must be centrally homogeneous. 
Now, the case (i) of the Proposition 1
demonstrates that the existence of a NS solution
depends on the solutions of a quartic equation.
For centrally homogeneous initial data these solutions exist
and are positive and real if and only if 
$\Theta_{u_0}\in \left]0, \Lambda_0^{3/2}(13/3-5\sqrt{3}/2)\right[\cup
\left]\Lambda_0^{3/2}(13/3+5\sqrt{3}/2), +\infty\right[$, with $\Lambda_0=M_3$.~~~$\Box$
\section{Conclusions} 
We have studied the final outcomes of the
inhomogeneous spherical dust collapse
as a function
of initial data, given in the form
(\ref{poly}) near the origin,
with the help
of families of radial null geodesics.
We have found all the possible cases where NS solutions
can arise in such collapse and
have demonstrated that assuming regular initial data then
the most general form of the equation (12)
is a quartic.
 
We have defined the notion of central homogeneity and
have proved that for the occurrence of
naked singularities the initial data
must in general be centrally homogeneous.
Mathematically this result indicates that
in general one would expect NS to be unstable
to centrally inhomogeneous
perturbations. It turns out, however,
that such perturbations are not physically
reasonable, in the sense that they would require
the density to tend to zero as
$r\to 0$, contrary to the physical expectation that
the density increases as we approach the centre.
 
In this way our results show that 
NS in the setting considered here remain 
robust when physically motivated perturbations
are allowed. They also provide us with an example
of the fact that stability and instability
of a particular phenomenon will crucially
depend on the class of perturbations allowed.
In this case, the occurrence of NS, though
mathematically unstable to general
centrally inhomogeneous perturbations, can
become stabilised once only
physically motivated perturbations (in the sense made precise above) are
allowed.

This is a potentially important point to bear in mind
in the general debates regarding the generic presence
of naked singularities in gravitational collapse.

It may also be noted that as far as the existence of naked singularities
and black holes as end product of collapse is concerned, rather general
results are available, with $M$ and $E$ being just $C^2$, without any further
restrictions (see e.g. \cite{Joshi-Dwivedi93}). However, these results deal with only
the existence part, and give no insight on possible distribution of these
outcomes. We believe this latter has been achieved here for the first time in
an explicit manner for a wide class of physically
reasonable initial data.

Finally we should add that most of our results are readily generalisable
to initial data characterised by the functional
form (\ref{Mgeneral}), see \cite{gen3}, 
and that generalizations to more generic settings than spherical 
dust are currently being investigated and will be the subject of a future publication.
\vspace{.2in}

\centerline{\bf Acknowledgments}

\vspace{.2in}
We thank Malcolm MacCallum and Brien Nolan for reading the manuscript.
FCM thanks Centro de Matem\'{a}tica, Universidade do Minho,
for support and FCT (Portugal) for grant PRAXIS XXI BD/16012/98.
RT benefited from PPARC UK Grant No. L39094. PSJ thanks the Astronomy
Unit, QMW, for hospitality and CERN for grant 
CERN/S/FAE/1172/97.

 

\begin{thebibliography}{}
\bibitem{Bondi} Bondi H, {\em Mon. Not. R. Astron. Soc.}, {\bf 107} (1947) 410

\bibitem{chris94} Christodoulou D, {\em Ann. of Math.}, {\bf 140} (1994) 607

\bibitem{chris99} Christodoulou D, {\em Ann. of Math.}, {\bf 149} (1999) 183

\bibitem{Deshingkar-etal99} Deshingkar S S, Joshi P S \& Dwidevi I H,
{\em Phys. Rev. D}, {\bf 59} (1999) 044018
 
\bibitem{Dwivedi-Joshi97} Dwivedi I H \& Joshi P S, 
{\em Class. Quant. Grav.}, {\bf 14} (1997) 1223

\bibitem{Hawking-Ellis73} Hawking S W \& Ellis G F R, {\em 
The large scale structure of space--time}, CUP (1985)

\bibitem{Jhingan-Joshi97} Jhingan S \& Joshi P S, {\em Ann. Isr. Phys. Soc.},
{\bf 13} (1997) 357 

\bibitem{Joshi-Book} Joshi P S, {\em Global aspects in gravitation 
and cosmology}, Oxford University Press (1993) 

\bibitem{Joshi-Dwivedi93} Joshi P S \& Dwivedi I H, {\em Phys. Rev. D}, 
{\bf 47} (1993) 5357  


\bibitem{Joshi-Dwivedi92} Joshi P S \& Dwivedi I H, {\em Commun. Math. Phys.},
{\bf 146} (1992) 333 

\bibitem{Krasinski} Krasi\'{n}ski, A., {\em Inhomogeneous cosmological models},
CUP, (1997) 

\bibitem{Lake92} Lake K, {\em Phys. Rev. Lett.}, {\bf 68} (1992) 3129 


\bibitem{Lemaitre} Lemaitre G, {\em Ann. Soc. Sci. Bruxelles}, {\bf
A53} 
(1933) 51 
 
\bibitem{Magli97} Magli G, {\em Class. Quant. Grav.}, {\bf 14} (1997)
1937 

\bibitem{Ori-Piran90} Ori A \& Piran T, {\em Phys. Rev. D}, {\bf 42}
(1990) 1068


\bibitem{Newman86} Newman  R P A C, {\em Class. Quant. Grav.}, {\bf
3} (1986) 527

\bibitem{Penrose} Penrose R, {\em Nuovo Cimento}, {\bf 1} (1969) 252

\bibitem{Tolman} Tolman R C, {\em Proc. Nat. Acad. Sci.}, {\bf 20}
(1934) 169

\bibitem{Wald97} Wald R, {\em preprint gr-qc/9710068} (1997)

\bibitem{rho-non-zero} Note that this form for $M$ also allows $\rho(0,0)=0$ but excludes e.g.
functions of the form $M(r)=e^{-1/r^k},~k>0,~for~r\ne0, 
~and~ M(r)=0,~for~r=0$, which are $C^\infty$ for all $r\in\Re$ but are non-analytic for any
open interval containing $r=0$. 

\bibitem{geo-scaling} For a general rescaling $R(0,r)=v(r)$, see \cite{gen}, it would be 
convenient to choose $u(r)=v^\alpha(r),~\alpha>0$.

\bibitem{gen} This is subject to the rescaling (6). A general
rescaling by $R(0,r)=v(r)$, where $v$ is 
a monotonically increasing positive real function of $r$ with $v(0)=0$,
would give the condition
$M(r)=O(v^a(r))$, around $r=0$, or equivalently $M(r)\le \beta v^a(r)$, around $r=0$,
where $\beta$ is a positive constant. 

\bibitem{homo} In fact for $M(r)^2=k E(r)^3$ there is always a rescaling $R(0,r)=M(r)^{1/3}$ 
such that $t_c(r)=const.$, independently of the particular forms of the functions
$M$ and $E$ considered.

\bibitem{gen2} In a number of comments,
 see \cite{gen,second-order,gen3}, we describe how our results can be generalized to more
generic forms for $M$ and $E$.

\bibitem{second-order} For a generalized form of the functions $M$ and $E$,
see \cite{gen3}, one should, at this point, consider
the generalized centrally homogeneous initial data: $M(r)=r^3g(r)+O(r^{i_1}),~i_1>3$, and 
$E(r)=r^2h(r)+O(r^{j_1}),~j_1>2$, where $g$ and $h$ are $C^2$ functions that remain non-zero 
and finite as $r\to0$.

\bibitem{gen3} All that is required in the
generalised proofs  
(except in the case of Lemma 2) is that $p_0$ is 
ensured to be finite by the functions $M(r)=O(r^a)$, as in (\ref{Mgeneral}), and
 $E(r)=O(r^b),~b\ge a-1$. Where $a$ and $b$ essentially
play the same role as $i_0$ and $j_0$ in the text. For example, one can consider 
non-analytic functions  
$M$ and $E$ which near the origin have the form
\bea
&&M(r)\approx M_{i_0} r^{i_0} + M_{i_1} r^{i_1},~~i_0,i_1\in [3,\infty),~
with~i_0<i_1,~M_{i_0}>0,~M_{i_1}\in \Re;\nonumber\\
&&E(r)\approx E_{j_0} r^{j_0} + E_{j_1} r^{j_1},~~j_0,j_1\in [0,\infty),~
with~j_0<j_1,~E_{j_0},~E_{i_1}\in \Re.\nonumber
\eea
Other examples of more general forms of $M$ were shown in \cite{rho-non-zero}.
 
Following the comments on 
\cite{gen} one should also add that a general rescaling $R(0,r)=v(r)$,
 would allow $M(r)=O(v^a(r))$ and $E(r)=O(v^b(r))$,~$b\ge a-1$, and our results follow
 accordingly. 
\end{thebibliography}
\end{document}